\documentstyle[11pt]{article}
\begin{document}
\vskip 22pt

\vspace{0.5cm}

\centerline{\large \bf  Empirical Calculations of the Formation
Energies of Intrinsic Defects in Lithium Niobate}
\vspace{1cm}

\centerline{Shaoxin Feng, Qinghua Jin, Baohui Li, Zhenya Guo,
Datong Ding}

\vspace{0.5cm}
{\small
\vspace{8pt}
\centerline{ Department of Physics, Nankai University, Tianjin 300071, P. R. China}
}

\date{\today}

\vspace{0.6cm}

\begin{center}
\begin{minipage}{11cm}

\noindent{\bf Abstract}

\vspace{8pt}

{\small

\noindent Results of our shell-model calculations concerning the
determination of empirical parameters of inter-ionic potentials
and the formation energies of point defects in LiNbO$_3$ are
presented. We employed the relaxed fitting method which is
particularly appropriate for the modeling of low-symmetry
structure. Attention is paid to guarantee the simulation precision
of the parameters who are sensitive for defect formation energy
calculations. It is shown that Li Frenkel disorders are the main
intrinsic defects in stoichiometric lithium niobate and Li vacancy
model is the dominant defect species in congruent lithium
niobate.}

\end{minipage}
\end{center}
\baselineskip 22pt

\vspace{0.5cm}

\section{Introduction}

\vspace{0.5cm}

Lithium Niobate (LiNbO$_3$) is widely used ferroelectric material
because of its favorable optical, piezoelectric, electro-optic,
elastic, photoelectric and photorefractive properties. LiNbO$_3$
has a tendency to non-stoichiometry with R=[Li]/[Nb]$<1$. Such
crystals therefore have very high concentration of intrinsic
defects.

The intrinsic defect structures of LiNbO$_3$ are widely
investigated because many physical properties of the crystal
depend to a certain extent on it. The required local charge
neutrality can theoretically be guaranteed by one of the models
listed in Table 1.

\begin{center}

Table 1. Notations of the intrinsic defect structure of LiNbO$_3$

\vspace{0.5cm}

\begin{tabular}{|c|c|c|c|}\hline
Model No. & Parametric formula & Kr\"{o}ger-Vink notation &
Authors\\ \hline 1 & Li$_{\rm 1-2x}$NbO$_{\rm 3-x}$ & 2V$_{\rm
Li}$V$_{\rm O}$ & Fay\cite{Fay}\\ \hline 2 & [Li$_{\rm
1-5x}$Nb$_{\rm x}$]NbO$_3$ & Nb$_{\rm Li}$4V$_{\rm Li}$ &
Lerner\cite{Lerner}\\ \hline 3 & [Li$_{\rm 1-5x}$Nb$_{\rm
5x}$]Nb$_{\rm 1-4x}$O$_3$ & 5Nb$_{\rm Li}$4V$_{\rm Nb}$ &
Abrahams\cite{Abrahams1} \\ \hline
\end{tabular}

\end{center}

Model 1 was suggested by Fay $et\; al$. (in 1968) that Li$_2$O
deficiency in LiNbO$_3$ is accommodated by the formation of Li and
O vacancies\cite{Fay}. Model 2 was suggested by Lerner $et\; al$.
(in 1968) that, in crystals with R=[Li]/[Nb]$<1$, excess Nb$^{5+}$
ions are substituted for Li$^+$ in their sites. To satisfy the
electro-neutrality condition, there will then arise lithium
vacancies. The Nb sublattice is full occupied and every four
vacant Li site (charge of -4) are statistically accompanied by one
Li site occupied by a Nb ion (difference charge of +4 at this
site)\cite{Lerner}. Model 3 was suggested by Abrahams and Marsh
(in 1986), that the O sublattice remains 100\%, where the
excessive charge of five Nb$^{+5}$ ions on Li-site (charge of +20)
is compensated by four vacant Nb-site (charge of
-20)\cite{Abrahams1}.

Although careful experimental investigation of the intrinsic
defect structure have brought some insight into a microscopic
understanding, there is no general agreement upon any one of the
defect model\cite{Donner1}.

Based on the shell model\cite{Dick}, Donnerberg $et\; al$. carried
out pioneer atomic simulation studies on defects in LiNbO$_3$
crystals\cite{Tom}\cite{Donner1}\cite{Donner2}, by which they can
then compare these defect structure models on the same energetic
analysis in terms of their formation energies.

In such empirical approach the model parameters as well as the
potential parameters are determined in terms of the empirical
parameterization, i.e. all the parameters involved are adjusted to
reproduce as closely as possible the measured crystal data.

Nevertheless, because of the low symmetry and the covalency
cohesion involved for the ternary oxide lithium niobate, the
practical limitations of the empirical parameterization bring
about that either the residual strain will be present, if the
crystal properties (cohesion, elastic constants, dielectric
constants and phonon spectrum) are fitted exactly, or conversely
the poor crystal properties may be obtained if all ion and bulk
strain is eliminated.

Since the fitting result is not unique, an optimum parameter set
has to be chosen deliberately according to succeeding researches.
Therefore, we refine upon some parameters which are sensitive for
defect formation energy calculations, whereas loosen the errors
for the other. Moreover, the fitting result is concerned with the
path of the optimization. In the present study, the relaxed
fitting approach by Gale\cite{Gale} is adopted.

Based on the optimum parameter set, the defect formation energies
can then be calculated by using the division strategy of Lidiard
and Norgett\cite{Lidiard}\cite{Norgett} incorporating with
Mott-Littleton approximation\cite{Mott}.

Lithium deficient non-stoichiometric LiNbO$_3$ can be imaged as
the result of Li$_2$O out-diffusion from stoichiometric one. There
are different out-diffusion modes assumed in respect to
corresponding defects emerging in the host LiNbO$_3$ after the
removal of the Li$_2$O. By comparing three diffusion modes in
terms of the energy needed for removing one Li$_2$O, one can
conclude which intrinsic defect model will be energetically
favorable in lithium deficient non-stoichiometric LiNbO$_3$.

\vspace{0.5cm}

\section{Parameterization}

\vspace{0.5cm}

According to the shell model, each ion is assumed to consist of an
core (charge $X$, mass$\rightarrow M_{ion}$) and an electron shell
(charge $Y$, mass$\rightarrow$0). Therefore the number of
constituent units of the crystal is doubled. The polarization of
the crystal is caused by the displacements of ions and the
relative displacements between cores and shells.

There are four kinds of interaction for the crystal.

(1) The interaction between shell and core belong to the same ion,
\begin{eqnarray}
\phi_{s-c} &=& {1\over 2} k_{sc} |{\bf r}_s-{\bf r}_c|^2
\end{eqnarray}
in which the spring constant $k_{sc}$ characterizes the harmonic
coupling between the shell and core.

(2) The Coulomb interactions between ions $i$ and $j$,
\begin{eqnarray}
\phi _{Coulomb} &=& {X_iX_j \over |{\bf r}_{ic}-{\bf
r}_{jc}|}+{X_iY_j \over |{\bf r}_{ic}-{\bf r}_{js}|}+{Y_iX_j \over
|{\bf r}_{is}-{\bf r}_{jc}|}+{Y_iY_j \over |{\bf r}_{is}-{\bf
r}_{js}|} \;\;\; i\neq j
\end{eqnarray}
where $i\neq j$ emphasizes that there is no Coulomb interaction
between shell and core belong to the same ions.

(3) The short-range interactions between shells are described with
Buckingham potential form:
\begin{eqnarray}
\phi _{non-Coulomb} (r) &=& A_{ij}\exp(r/\rho_{ij})-C_{ij}/r^6
\end{eqnarray}
where $r$ is the distance between the centers of electron shells
of species $i$ and $j$. Because the radius of cation is relatively
small the interaction between cations is ignored. Besides, the van
der Waals terms $-C_{ij}/r^6$ in anion-cation interaction is
omitted too.

(4) Three body bond-bending terms are used to model the
directional character of covalent bond:
\begin{eqnarray}
\phi _{bend} (\theta) &=& {1 \over 2} k_B (\theta-\theta_0)^2
\end{eqnarray}
where $k_B$ is the bond-bending force constant, $\theta$ is the
bond angle of O-Nb-O, and $\theta_0$ the equilibrium bond angle.

Accordingly to the equations listed above, there are 14 parameters
involved. Since the polarizability of Li ion, $\alpha _{\rm
Li}=Y_{\rm Li}^2 / k_{sc}^{\rm Li}$ , is very small, so we take
approximately $k_{sc}^{\rm Li}\rightarrow \infty$, $Y_{\rm Li}=0$
. Therefore the member of undetermined parameters reduced to 12. A
parameter set can be regarded as a point in the "12-dimensional
parameter space".

We can properly choose an initial parameter set as a starting
point and then move the point closed to the reasonable position by
an optimization procedure. In the present study, "relaxed fitting"
approach by Gale is employed. This approach based on displacements
of the energy-minimized configuration from the experimental
structure is found to be superior to conventional fitting in which
zero gradients of the experimental geometry is the aim. The
relaxed fitting yields the exact displacements and genuine
physical properties and appears to be particularly appropriate for
the modeling of low-symmetry structure\cite{Gale}.

In order to quantify the degree of the optimization, we define an
objective function as following:
\begin{eqnarray}
F &=& u\sum\limits _i
(S_{iCalc}-S_{iExpr})^2/S_{iRef}^2+v\sum\limits_j
(P_{jCalc}-P_{jExpr})^2/P_{jRef}^2 \;\; u,v>0
\end{eqnarray}

It is obviously that $F>0$ and small $F$ implies small fitting
errors or high degree optimization. $\{ S\}$ and $\{ P\}$ denote
the quantities of crystal structure and property respectively. $u$
and $v$ are weighting factors. Generally, the precision of
reproduced structure should be more emphasized in comparing with
that of reproduced properties. Therefore, in the present study, we
take $u=100$ and $v=1$.

$(S_{iCalc}-S_{iExpr})^2/S_{iRef}^2$ and
$(P_{jCalc}-P_{jExpr})^2/P_{jRef}^2$ represent the fitting errors
between that calculations and experiment date. In order to avoid
the irrationality resulting from small denominator, we take the
referential values instead of the experiment data in some cases.
For instance, the fractional coordinate y(=0.3446) of the oxygen
ion in hexagonal cell is many times larger than x(=0.0492) and
z(=0.0647). But in practice, we take their referential values as
x$_{Ref}$=y$_{Ref}$=z$_{Ref}$=1.0. Similar treatments are adopted
for elastic constants(we use 10$\times 10^{10}$N m$^{-2}$ as their
referential values).

The object function, $F$, is then minimized in terms of
Davidon-Fletcher-Powell variable-metric algorithm. Since the
optimization often gets stuck in local minimum, and so we have to
try hundreds of initial parameter sets and then choose the globe
minimum from hundreds of local minima.

\begin{center}

Table 2. The two parameter sets of ionic interaction and shell
model

\vspace{0.5cm}

\doublerulesep 1.pt
\begin{tabular}{|p{1.4in}|c|c||p{1.4in}|c|c|}
\hline Parameters of short-range interaction & Set I & Set II &
Parameters of shell model & Set I & Set II \\ \hline $A_{\rm
Nb-O}$/ev & 5121.9 & 6373.6 & $Y_{\rm Nb}/|e|$ & -36.0 & -36.0 \\
\hline $\rho _{\rm Nb-O}/$\AA & 0.27543 & 0.27211 & $k_{sc}^{\rm
Nb}/$(ev\AA$^{-2})$ & 16553. & 13374. \\ \hline $A_{\rm Li-O}$/ev
& 144.27 & 246.64 & $Y_{\rm O}/|e|$ & -2.5376 & -2.4660 \\ \hline
$\rho_{\rm Li-O}$ /\AA & 0.35614 & 0.32554 & $k_{sc}^{\rm
O}$/(ev\AA $^{-2}$) & 11.397 & 11.459 \\ \hline \cline{4-6}
$A_{\rm O-O}$/ev & 270.35 & 198.80 & Parameters of bond-bending &
Set I & Set II \\ \hline $\rho_{\rm O-O}$/\AA & 0.40510 & 0.42612
& $k_B$/(ev rad$^{-2}$) & 0.0 & 0.51548 \\ \hline $C_{\rm
O-O}/$ev\AA$^6$) & 11.117 & 21.277 & $\theta_0/^\circ$ & | & 90
\\ \hline
\end{tabular}

\end{center}

Table 2 lists two parameter sets obtained, their corresponding
fitting results upon crystal structure and crystal property are
listed in Table 3 and 4 respectively.

\begin{center}
Table 3: Comparison between experimental and calculated values of
the structure of LiNbO$_3$ crystal in hexagonal cell.

\vspace{0.5cm}
\begin{tabular}{|c|c|c|c|}
\hline Structure & Exptl.\cite{Abrahams2} & Calcd. by set I &
Calcd. by set II \\ \hline a & 5.148 & 5.1477 & 5.2547 \\ \hline c
& 13.863 & 13.564 & 13.906 \\ \hline x$_{{\rm O}c}$ & 0.0492 &
0.01481 & 0.01816 \\ \hline x$_{{\rm O}s}$ & 0.0492 & 0.06293 &
0.06713 \\ \hline y$_{{\rm O}c}$ & 0.3446 & 0.37173 & 0.35736 \\
\hline y$_{{\rm O}s}$ & 0.3446 & 0.34061 & 0.36516 \\ \hline
z$_{{\rm O}c}$ & 0.0647 & 0.06385 & 0.07553  \\ \hline z$_{{\rm
O}s}$ & 0.0647 & 0.07164 & 0.08350 \\ \hline z$_{{\rm Nb}s}$ & 0.0
& -0.00026 & 0.00100 \\ \hline z$_{\rm Li}$ & 0.2829 & 0.28738 &
0.27500 \\ \hline
\end{tabular}
\begin{minipage}{13cm}
{\small where a and c are lattice constants. (x, y, z) are
fractional coordinates. The subscript $c$ and $s$ represent ion
core and electron shell respectively. We consider the core and the
shell are not coincide when crystal is under equilibrium status,
$i.\; e.$ there is a net dipole within the unit cell. The core and
shell of Nb ion are in the triadaxis with coordinates (0, 0, 0)
and (0, 0, z$_{{\rm Nb}s}$) respectively. Because the polarization
of Li ion is omitted its core and shell has the same coordinate
(0, 0, z$_{\rm Li}$).}
\end{minipage}

\vspace{0.5cm}

Table 4: Comparison of experimental and calculated data of crystal
properties for LiNbO$_3$

\vspace{0.5cm}
\begin{tabular}{|c|c|c|c|c|}
\hline Crystal & Exptl.\cite{Weis} & Calcd. of & Calcd. of &
Calcd. of  \\ properties & & Tomlinson & our set I & our set II \\
\hline C$_{11}$ & 20.3 & 25.7 & 21.30 & 21.27 \\\hline C$_{12}$ &
5.3 & 12.5 & 8.91 & 5.14 \\\hline C$_{13}$ & 7.5 & 10.6 & 6.90 &
7.47
\\\hline C$_{14}$ & 0.9 & -2.2 & -3.67 &  2.66 \\\hline C$_{33}$ &
24.5 & 27.6 & 22.28 & 25.19 \\\hline C$_{44}$ & 6.0 & 8.8 & 6.92 &
6.86
\\\hline C$_{66}$ & 7.5 & 6.6 & 6.20 & 8.06 \\\hline $\varepsilon_{11}(0)$
& 84.1 & 29.7 & 81.07 & 27.48 \\\hline $\varepsilon_{33}(0)$ &
28.1 & 61.8 & 33.56 & 40.74 \\\hline $\varepsilon_{11}(\infty)$ &
46.5 & 2.06* & 30.45 & 17.68 \\\hline $\varepsilon_{33}(\infty)$ &
27.3 & 1.83* & 21.35 & 25.68 \\\hline
\end{tabular}
\begin{minipage}{13cm}
{\small where $\varepsilon_{11}(0)$, $\varepsilon_{33}(0)$ are low
frequency dielectric constants, $\varepsilon_{11}(\infty)$,
$\varepsilon_{33}(\infty)$ are high frequency dielectric
constants, C$_{11}$, C$_{12}$, C$_{13}$, C$_{14}$, C$_{33}$,
C$_{44}$, C$_{66}$ $(10^{10} N m^{-2})$ elastic constants.}
\end{minipage}
\end{center}

It seems that the fitting results of the crystal properties result
from set I are better than that of set II. However the parameters
set I leads to imaginary modes of phonons, which indicates
structural unstable. Therefore parameters set I is given up and
parameter set II is the accepted result to be used for intrinsic
defect studies.

In comparing with the results of Tomlinson $et\; al$.\cite{Tom}
(see Table 4), two of the "most noticeable deviations", as
indicated by Tomlinson $et\; al$., namely the static dielectric
constant $\varepsilon_{33}(0)$ and the elastic constant C$_{14}$
are reduced in the present study. Moreover, according to the
assessment of us, the calculated high frequency dielectric
constant $\varepsilon_{11}(\infty)$ and $\varepsilon_{33}(\infty)$
results from the parameter set reported by Tomlinson $et\; al$.
would be 2.06 and 1.83(marked with '*' in table 4), such
deviations with magnitudes in order, are diminished considerablely
as well in our study.

\vspace{0.5cm}

\section{Lattice formation energy and point defect formation energy}

\vspace{0.5cm}

The lattice formation energy of LiNbO$_3$ is deduced from
parameter set II listed in Table 5. Then we extrapolate parameter
set II to the empirical calculation for the lattice formation
energies of Li$_2$O and Nb$_2$O$_5$ as well. The results obtained
are also listed in Table 5.

By using the division strategy of Lidiard and Norgett
incorporating with Mott-Littleton approximation, we can then
calculated the point defects in LiNbO$_3$. The formation energies
of different point defects listed in Table 5 are calculated by
using an improved homemade program similar to
HADES\cite{Norgett}\cite{Catlow}. The results are then combined to
give Frankel and Schottky energies which are also represented as
the total energy and the energy per defect (shown in parenthesis,
i.e. the Frenkel-pair energy divided by two, the energy of the
Schottky quintet in LiNbO$_3$ divided by five, the energy of the
Schottky trio in Li$_2$O divided by three and the energy of the
Schottky heptad in Nb$_2$O$_5$ divided by seven), and are shown in
Table 6.

\begin{center}

Table 5: Some typical lattice energies and defect energies

\vspace{0.5cm}

\begin{tabular}{|c|c|c|r|}
\hline Lattice & LiNbO$_3$ & E(LiNbO$_3)_{unit \; cell}$ & -183.95
ev \\ \cline{2-4} energies & Li$_2$O & E(Li$_2$O)$_{\rm unit \;
cell}$ & -31.67 ev \\ \cline{2-4}  & Nb$_2$O$_5$ &
E(Nb$_2$O$_5)_{unit \; cell}$ & -328.10 ev \\ \hline  & Li vacancy
& E(V$'_{\rm Li})$ & 8.73 ev \\ \cline {2-4}  & Li interstitial
site & E(Li$_I^\bullet)$ & -6.76 ev \\ \cline{2-4} Formation & Nb
vacancy & E(V$_{\rm Nb}^{5'})$ & 129.46 ev \\ \cline{2-4} energies
& Nb interstitial site & E(Nb$_I^{5\bullet})$ & -119.15 ev \\
\cline{2-4} of defect & O vacancy & E(V$_{\rm
O}^{\bullet\bullet})$ & 18.83 ev \\ \cline{2-4}  & O interstitial
site & E(O$''_I)$ & -10.16 ev \\ \cline{2-4}  & Nb antisite &
E(Nb$_{\rm Li}^{4\bullet})$ & -114.82 ev \\ \hline
\end{tabular}

\vspace{0.5cm}

Table 6 The formation energies of Frenkel and Schottky defects in
LiNbO$_3$ crystal.

\vspace{0.5cm}

\begin{tabular}{|l|r|}
\hline Li Frenkel pair & 1.97 (0.98) ev \\ \hline Nb Frenkel pair
& 10.31 (5.15) ev \\ \hline O Frenkel pair & 8.67 (4.33) ev \\
\hline LiNbO$_3$ Schottky  & 10.73 (2.15) ev \\ \hline Li$_2$O
Schottky & 4.62 (1.54) ev \\ \hline Nb$_2$O$_5$ Schottky & 24.97
(3.57) ev \\ \hline
\end{tabular}
\vspace{0.5cm}
\end{center}

\section{Intrinsic defect structure in Li deficiency LiNbO$_3$}

\vspace{0.5cm}

Li$_2$O deficiency non-stoichiometric LiNbO$_3$ could be imaged as
the result of Li$_2$O out-diffusion from stoichiometric LiNbO$_3$.
Theoretically there are three Li$_2$O out-diffusion modes related
to three intrinsic defect models respectively, namely:
\begin{eqnarray}
'{\rm LiNbO}_3'\longleftrightarrow {\rm Li}_2{\rm O+2V}'_{\rm
Li}+{\rm V}_{\rm O}^{\bullet\bullet}\;\;\;\;\;(\rm model\;1)
\end{eqnarray}
\begin{eqnarray}
'{\rm LiNbO}_3'\longleftrightarrow {\rm 3Li}_2{\rm O+4V}'_{\rm
Li}+{\rm Nb}_{\rm Li}^{4\bullet}\;\;\; (\rm model\;2)
\end{eqnarray}
\begin{eqnarray}
'{\rm LiNbO}_3'\longleftrightarrow {\rm 3Li}_2{\rm O+4V}^{5'}_{\rm
Nb}+{\rm 5Nb}_{\rm Li}^{4\bullet}\;\; (\rm model\;3)
\end{eqnarray}

The niobium vacancy (V$_{\rm Nb}^{5'}$) with effective charge of
$-5|e|$ tends towards to the niobium antisite (Nb$_{\rm
Li}^{4\bullet}$) with effective charge of $+4|e|$, which decreases
their total formation energy. Table 7 illustrates the formation
energy dependence of the distance between V$_{\rm Nb}^{5'}$ and
Nb$_{\rm Li}^{4\bullet}$. It can be seen that the nearest
neighboring V$_{\rm Nb}^{5'}$-Nb$_{\rm Li}^{4\bullet}$ pair,
complex defect [V$_{\rm Nb}^{5'}$Nb$_{\rm Li}^{4\bullet}$], has
the lowest formation energy. Therefore, a slight modification of
(8) yields
\begin{eqnarray}
'{\rm LiNbO}_3'\longleftrightarrow 3{\rm Li}_2{\rm
O+4(V}^{5'}_{\rm Nb}{\rm Nb}_{\rm Li}^{4\bullet})'+{\rm Nb}_{\rm
Li}^{4\bullet}
\end{eqnarray}

\begin{center}
Table 7: The relation between V$_{\rm Nb}^{5'}$|Nb$_{\rm
Li}^{4\bullet}$ distance and formation energy.

\begin{tabular}{|c|c|c|c|c|}
\hline & & next-next & next & \\ V$_{\rm Nb}^{5'}$|Nb$_{\rm
Li}^{4\bullet}$ distance & infinite & nearest & nearest & nearest
\\ & & neighbor & neighbor & neighbor
\\ \hline total formation energy & 14.64ev & 12.73ev & 12.29ev &
9.98ev \\ \hline
\end{tabular}
\end{center}

Table 8 contains our calculation results in which all the energies
needed for (6)$\sim$(9) out-diffusion modes are converted to per
Li$_2$O. In order to make a comparison with the previous studies,
the corresponding results reported by Donnerberg et
al.\cite{Donner1} are listed in Table 8 as well.

\begin{center}

Table 8 The defect energies per Li$_2$O of four typical defect
model.

\vspace{0.5cm}
\begin{tabular}{|c|c|c|c|c|}
\hline Process of out-diffusion & Eq. (6) & Eq. (7) & Eq. (8) &
Eq. (9) \\ \hline Calcd. by us & 4.62ev & 3.01ev & 10.89ev &
4.68ev \\ \hline Calcd. by Donnerberg & 5.82ev & 4.56ev & 15.2 ev
& 10.7ev \\ \hline
\end{tabular}
\end{center}

The energy difference between Eq.(6) and Eq.(7) indicates the
model 1 (oxygen vacancy model) is energetically unfavorable.
Though the energy denoted by Eq.(9) is much less than
corresponding result reported by Donnerberg $et\; al$.[4], it is
still 1.5 times larger than the energy denoted by Eq.(7).

The fact that the energy denoted by Eq.(7) is the lowest implys
the defect structure model 2 (lithium vacancy model) is
energetically favorable.

\vspace{0.5cm}

\section{Conclusion and discussion}

\vspace{0.5cm}

The simple defect model 1 was rejected by several experimental
evidences. Until 1993 the model 3 was favored. However, the
earlier NMR studies\cite{Peterson} which had strongly supported
the niobium vacancy model is in doubt\cite{Blumel}.

The x-ray single-crystal diffraction and the TOF neutron powder
diffraction investigation by Iyi $et\; al$. strongly supported the
Li-site vacancy model\cite{Iyi}. Wilkinson $et\; al$. examined the
congruently melting lithium niobate samples by powder x-ray
diffraction techniques and found the Nb site is fully
occupied\cite{Wilkinson}. The x-ray and neutron powder diffraction
experiments by Zotov $et\; al$. shown that the Li-sites vacancy
model describes best the average defect structure in congruent
lithium niobate\cite{Zotov}.

Recently, by using a formula for calculating the soft mode
frequency and Curie temperature to the analysis of the defect
structure, Safaryan $et\; al$. concluded that the lithium vacancy
model best describes the defect structure of nonstoichiometric
lithium niobate\cite{Safaryan}.

According to the present empirical study, we can conclude that
lithium deficiency of the LiNbO$_3$ is accompanied mainly with the
formation of lithium vacancies and niobium antisite.

In comparing with the result reported by Tomlinson $et\; al$., the
degree of fit between the calculated crystal properties and the
observed data has been improved in the present study, especially
the high frequency dielectric constants. Even though there exist
still obviously the disagreement between
$\varepsilon_{11}(\infty)$ calculated and experimental values, it
will take less effect for the polarization behavior of the ion
under the electric displacement ($D$), which is important for
successive studies of intrinsic defects. The reason is that the
polarization density $P$ can be expressed as $P={\varepsilon-1
\over \varepsilon}D$ (deduced from
$D=\varepsilon_0E+P=\varepsilon_0\varepsilon E$), and so the
proportionality factors are 0.944 and 0.978 corresponding to
$\varepsilon_{11}(\infty)$ calculated value of 17.1 and
experimental value of 46.5 respectively.

\vspace{0.7cm}

\noindent{\bf Acknowledgment:}

This work is supported by the National High Technology Development
Program of China (Grant No. 715-001-0102).

\end{document}